\documentclass[prl,twocolumn,showpacs,preprintnumbers,amssymb,superscriptaddress,nofootinbib]{revtex4-1}
\usepackage{braket} 
\usepackage{graphicx}
\usepackage{amsmath}
\usepackage{bm}
\usepackage[usenames]{color}
\usepackage{color}

\def\>{{\rangle}}
\def\<{{\langle}}
\def\)>{{)\!\rangle}}
\def\(<{{\langle\!(}}

\graphicspath{{../figs/}}

\begin{document}

\title{Dynamics of high-harmonic generation in terms of complex Floquet spectral analysis}

\author{Hidemasa \surname{Yamane}}
\email{s\_h.yamane@p.s.osakafu-u.ac.jp}
\affiliation{Department of Physical Science, Osaka Prefecture University, Gakuen-cho 1-1, Sakai 599-8531, Japan}

\author{Ken-ichi \surname{Noba}}
\affiliation{Department of Physical Science, Osaka Prefecture University, Gakuen-cho 1-1, Sakai 599-8531, Japan}

\author{Tomio \surname{Petrosky}}
\affiliation{Center for Complex Quantum Systems, The University of Texas at Austin, Austin, TX 78712 USA}
\affiliation{Institute of Industrial Science, University of Tokyo, 4-6-1 Komaba, Meguro,
Tokyo 153-8505, Japan}

\author{Satoshi \surname{Tanaka}}
\email{stanaka@p.s.osakafu-u.ac.jp}
\affiliation{Department of Physical Science, Osaka Prefecture University, Gakuen-cho 1-1, Sakai 599-8531, Japan}

\begin{abstract}
High-harmonic generation (HHG) of a two-level-system driven by an intense monochromatic phase-locked laser is studied in terms of complex spectral analysis with the Floquet method.
In contrast to the phenomenological approaches, this analysis deals with the whole process as a coherent quantum process based on microscopic dynamics.
The spectral decomposition corresponding to the contributions of the Floquet resonance and dressed continuous states of the total system have been obtained.
The calculated HHG spectrum exhibits the characteristic features of the HHG from solids.
We found that the quantum interference of the Floquet resonance states is responsible for the transition from the adiabatic to the stationary regime in the HHG process and
that the phase of the driving laser controls the dynamics of the HHG photon emission.
\end{abstract}
\date{\today}

\maketitle


The recent development of ultra-intense pulse lasers facilitates high-harmonic generation (HHG) of wavelengths in the X-ray region \cite{Corkum93PRL,Brabec00RMP,Krausz09RMP}.
Furthermore, the precise control of the carrier-envelop phase (CEP) of the driving laser enables the capture and quantum mechanical control of ultrafast electronic wavepacket motions in the attosecond timescale in atoms and molecules\cite{Krausz09RMP}.

HHG from solids has been experimentally observed recently, where the HHG spectrum is found to be characteristically different from those with ionized atomic gas \cite{Ghimire11NatPhys,Schubert14NaturePhoto,Hohenleutner15Nature}.
For example, the cutoff energy of HHG is found to be proportional to the laser amplitude, in contrast to the case with atomic gases where it is proportional to the laser intensity. 
Consequently, the HHG mechanism from solids is considered to be fundamentally different from that with atomic gas, which has been interpreted by the so-called {\it three-step model} \cite{Corkum93PRL,Brabec00RMP,Krausz09RMP}.  
Now many debates have intensely continued regarding the microscopic mechanism of the HHG process \cite{Ghimire11NatPhys,Schubert14NaturePhoto,Hohenleutner15Nature,Vampa15Nature,Wu16PRA}.

In most of the conventional theories of HHG, the coherent electronic motion under the strong driving field is first calculated, and then the HHG spectrum is obtained for the given electronic current or polarization \cite{Corkum93PRL,Brabec00RMP,Krausz09RMP,Ghimire11NatPhys,Schubert14NaturePhoto,Hohenleutner15Nature,Vampa15Nature,Wu16PRA}. 
This treatment separates the HHG photon emission process from the coherent excitation of the electron, hence, the quantum correlation between  matter and the radiation field might be lost. 
Instead, it is necessary to deal with the whole process, from the electronic excitation to the HHG photon emission, as a coherent quantum process based on microscopic dynamics.

It is then a challenging task to describe the irreversible dynamics of a driven open quantum system within the ordinary realm of quantum mechanics, 
since an irreversible decay process cannot take place because the Hermitian Hamiltonian takes only real eigenvalues within the Hilbert vector space. 
As a solution to the problem, a new formalism, i.e., {\it complex spectral analysis}, has been explored over the last two decades so that the Hamiltonian can take complex eigenvalues by expanding the vector space to the {\it extended Hilbert space}\cite{Petrosky91Physica,Petrosky97AdvChem,Ordonez01PRA,Petrosky01PRA,Tanaka06PRB,Yamada12PRB,Tanaka16PRA,Fukuta17PRA}.
 
In this Rapid Communication, we apply the complex spectral analysis to study the HHG, using the Floquet method to take into account a non-perturbative interaction between the matter and the driving field. 
The total system under consideration consists of not only the strongly coupled  matter and driving laser system but also the scattered radiation field with infinite degrees of freedom.
We solve  the complex eigenvalue problem of the Floquet Hamiltonian of the total system in the extended Hilbert space. 
The time evolution of the state vector of the total system is then clarified by the complex spectral expansion.
In this Communication, some results are presented on the HHG of a two-level-system (TLS) driven by an intense monochromatic phase-locked laser.

The calculated HHG spectrum exhibits the characteristic features of the HHG from solids.
We reveal that the quantum interference of the Floquet resonance states is responsible for the transition from the adiabatic to the stationary regime in the HHG process
and the phase of the driving laser controls the dynamics of the HHG photon emission.

%
%
%

\begin{figure}
\includegraphics[width=70mm]{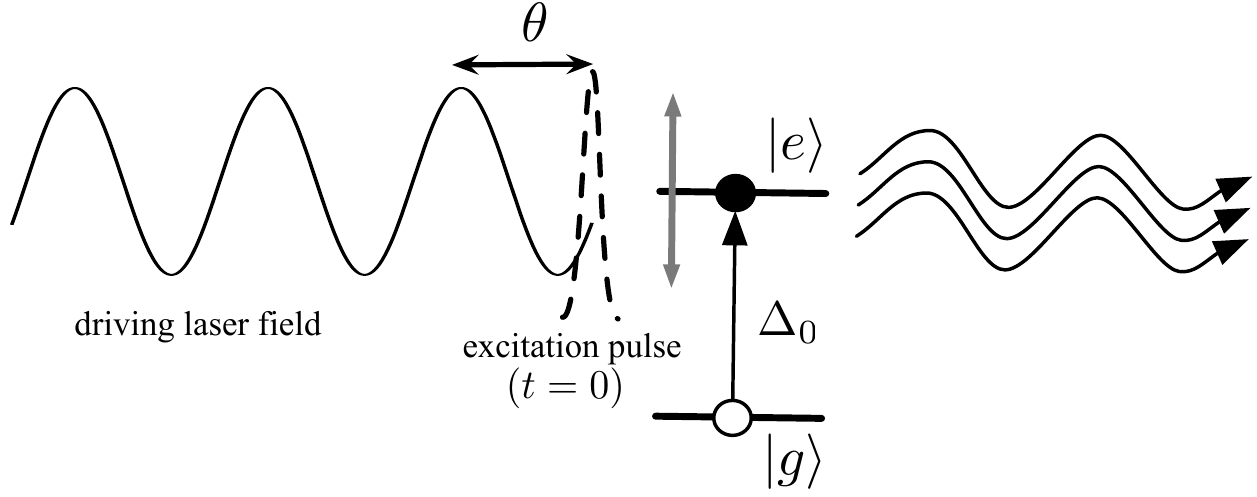}
\caption{High-harmonic generation of the driven TLS.}
\label{fig:model}
\end{figure}

{\it Model.}
The spontaneous emission from a TLS composed of the ground state $|g\>$ and  the excited state $|e\>$ with the energies $E_0$ and $E_e$, respectively is considered, as shown in Fig.\ref{fig:model}.
The excitation energy $E_e$ is periodically changed by a monochromatic phase-locked laser field with the amplitude $A$, frequency $\omega$ and the initial phase $\theta$.
Under the dipole approximation and rotating wave approximation, the minimal coupling Hamiltonian is given by 
\begin{align}\label{eq:Ht1}
\hat H(t)&=E_0|g\>\<g| +\left (E_e+A\cos(\omega t +\theta)\right)|e\>\<e|
\notag\\
&+ \int \omega_{k} \hat a_{k}^\dagger \hat a_k  dk+\lambda \int {\cal C}_{ k} \left( |e\>\<g| \hat a_{ k} + |g\>\<e| \hat a_{ k}^\dagger \right) dk \;,
\end{align}
where  $\hat a_{k}$ ($\hat a^\dagger_{ k}$) represents the scattered radiation field with energy $\omega_k=c|k|$, and $\lambda$ is a dimensionless coupling constant, where we consider $c=1$.
The coupling coefficient ${\cal C}_k$ is given by ${\cal C}_k=\sqrt\omega_k$ except for a constant factor\cite{MilonniBook}.
In Eq.(\ref{eq:Ht1}), we regard the driving laser field as a classical field, while we treat the scattered radiation field  quantum mechanically.
In this Rapid Communication, we consider a one-dimensional system, where the discrete variable of the wave number becomes a continuous variable in the large size limit.
Since the number of elementary excitations does not change in $\hat H(t)$,  a single-excitation subspace composed of the dressed atom states of $|e\>\otimes |0_k\>\equiv |d\>$ and $|g\>\otimes|1_k\>\equiv |k\>$ is considered  \cite{Cohen98atomBook}.
Then the Hamiltonian $\hat H(t)$ in this subspace is represented by
$\hat{H}(t)=\left(\Delta_0+A\cos(\omega t+\theta) \right)|d\>\<d|+\int \omega_{k}| k\>\< k| dk+ \lambda\int{\cal C}_{ k}(|d\>\< k|+| k\>\<d|) dk ,
$
where $\Delta_0\equiv E_e-E_0$, and we take $E_0$  as the origin of energy.

Since the Hamiltonian is time-periodic, $\hat{H}(t+T)=\hat{H}(t)$ with $T\equiv2\pi/\omega$, the Floquet theorem may be applied: The wave vector can be written as
$
|\Psi(t)\>=\sum_\xi c_\xi e^{-iz_\xi t}|\Phi_\xi (t)\>,
$
with a periodic Floquet eigenfunction 
$|\Phi_\xi(t+T)\>=|\Phi_\xi(t)\>$ \cite{Grifoni98PhysRep,Shirley65PR,Sambe73PRA}.
The composite space ${\cal F}\equiv {\cal R}\otimes {\cal T}$ is made up of the configuration space $\cal R$ and the  space $\cal T$ of functions which are periodic in time with period $T$ \cite{Grifoni98PhysRep}. 
 The conjugate basis set $\{|\kappa_n)\}$ to the time basis set $\{|t)\}$ is constructed as
$|\kappa_n)\equiv {1\over T}\int_0^Tdt e^{i\kappa_n t}|t)$, 
where
$\kappa_n\equiv n\omega=2\pi n/T\quad (n=0,1,\cdots)$ \cite{Yamane18PIERS}.
 
In terms of the conjugate basis set, the Floquet Hamiltonian is represented by
$
\hat{H}_{\rm F}=\sum_{n=-\infty}^{\infty}(\Delta_0+n\omega)|d,\kappa_n\)>\(<d,\kappa_n| 
+\sum_{n=-\infty}^{\infty}(A/2) [ e^{i\theta} |d,\kappa_{n+1}\)>\(<d,\kappa_n|+ e^{-i\theta} |d,\kappa_n\)>\(<d,\kappa_{n+1}| ] 
+\sum_{n=-\infty}^{\infty}\int (\omega_{k}+n\omega)| k,\kappa_n\)>\(< k,\kappa_n| dk 
+ \lambda\sum_{n=-\infty}^{\infty}\int   {\cal C}_{k} [ | k,\kappa_n\)>\(<d,\kappa_n|+|d,\kappa_n\)>\(< k,\kappa_n|  ] dk \;,
$
where $|\cdot)\!\>$ denotes the vector in the composite space $\cal F$.
With the use of the Stark basis defined by 
$
|\phi_{d}^{(n)}\)>=\sum_{m=-\infty}^{\infty}e^{-i(n-m)\theta} J_{n-m}(a)|d,\kappa_m\)> 
$, where $J_{n}(x)$ is the $n$-th order Bessel function of the first kind and $a\equiv A/\omega$, 
the Floquet Hamiltonian $\hat{H}_{\rm F}$ is rewritten as
\begin{align}
\label{eq:FH2}
\hat{H}_{\rm F}&=\sum_{n=-\infty}^{\infty}\left\{ (\Delta_0+n\omega) |\phi_{d}^{(n)}\)>\(<\phi_{d}^{(n)}| \right. \notag\\
&\left.  \hspace{1cm} + \int (\omega_{k}+n\omega)| k,\kappa_n\)>\(< k,\kappa_n|  \right\} dk \nonumber\\
&+\lambda\sum_{n,m=-\infty}^{\infty}\int  {\cal C}_{ k}J_{m-n}(a) \left( e^{-i(m-n) \theta} | k,\kappa_n\)>\(<\phi_{d}^{(m)}| \right. \notag\\
&\left.  \hspace{2cm} + e^{i(m-n)\theta}|\phi_{d}^{(m)} \)>\(< k,\kappa_n|  \right) dk \;.
\end{align}

{\it Complex spectral analysis.} 
The right and left eigenvalue problems of $\hat{H}_{\rm F}$ are given by
\begin{align}
\label{eq:rightev}
\hat{H}_{\rm F}|\Phi_\xi^{(n)}\)>=z_\xi^{(n)}|\Phi_\xi^{(n)} \)>,\quad \(<\widetilde{\Phi}_\xi^{(n)}|\hat{H}_{\rm F}=z_\xi^{(n)}\(<\widetilde{\Phi}_\xi^{(n)}|\;,
\end{align}
where $|\Phi_\xi^{(n)} \)>$ and $\(<\widetilde{\Phi}_\xi^{(n)}|$ are the right and left eigenstates with the same complex eigenvalue $z_\xi^{(n)}$.
It is well known that the Floquet eigenstate possesses mode-translational symmetry $|\Phi_\xi^{(n)}(t)\>=\exp[-i \kappa_n t ] | \Phi_\xi^{(0)}(t) \>$ with the shifted quasienergy $z_\xi^{(n)}=z_\xi^{(0)}-n\omega$
  \cite{Grifoni98PhysRep}.
The eigenstates satisfy the bi-completeness and  bi-orthonormal relation \cite{Petrosky91Physica,Yamada12PRB,Yamane18PIERS}.
The eigenvalue problem of the Floqeut Hamiltonian was solved in terms of the Brillouin-Wigner-Feshbach projection method\cite{Tanaka16PRA,Fukuta17PRA}.
The detailed derivation is shown in Ref.\cite{Yamada12PRB,Yamane18PIERS}.

The right-resonance eigenstate in the weak coupling case $\lambda\ll1$ is obtained by 
\begin{align}\label{PhidRL}
|\Phi_d^{(n)}\)>&={\cal N}_d^{(n)}\left\{ |\phi_{d}^{(n)}\)>  \right.  \notag\\
 &  \left. +\lambda\sum_m \int dk  {\cal C}_k  { J_{n-m}(a)e^{-i(n- m) \theta} \over [  z-(\omega_k+m\omega)]^+_{z=z_{d}^{(n)}}}|{k},\kappa_{m}\)>\right\}\;,
\end{align}
where the $+$ sign in the denominator of Eq.(\ref{PhidRL}) indicates taking the analytic continuation from the upper half of the complex energy plane\cite{Petrosky91Physica}, and ${\cal N}_d^{(n)}$ is a normalization constant.
The left-resonance eigenstates are also obtained by first taking the Hermite conjugate, and then the same analytic continuation instead of the opposite analytic continuation 
\cite{Petrosky91Physica,Yamada12PRB,Yamane18PIERS}. 
The complex eigenvalue of the resonance state is given by
\begin{align}
\label{eq:dev}
 z_{\tilde d}^{(n)}=\Delta_0+n\omega + \lambda^{2}\sum_{n=-\infty}^{+\infty}\sigma^+(\Delta_0+(n-m)\omega)J_{n-m}^{2}(a)\;,
\end{align}
where the scalar self-energy function is given by $\sigma^+(z)\equiv \int  {\cal C}_k^2/[z-\omega_k]^+dk$.
Note that the resonance state decays exponentially with the decay rate given by the imaginary part of $z_{\tilde d}^{(n)}$.

The dressed continuous right-eigenstates are also obtained by
\begin{align}\label{eq:Phik}
&|\Phi_{k}^{(n)}\)> \notag\\
&=|k,\kappa_n\)>+\lambda {\cal C}_k \sum_{m}{ J_{m-n}(a)e^{i(m-n)\theta} \over \omega_k+n\omega+i0^+-{\cal A}_{m,DL}^+(\omega_k+n\omega) } \notag \\
&\times\left\{ |\phi_d^{(m)} )\!\>   + \lambda\sum_{m'}\int {  {\cal C}_{k'} J_{m-m'}(a) e^{-i(m-m')\theta} | k',\kappa_{m'} \)>dk' \over  \omega_k-\omega_{k'}+(n-m')\omega+i0^+ }  \right\}
\end{align}
where ${\cal A}_m^+(\varepsilon)\equiv\Delta_0+n\omega+\lambda^2\sum_m J_{n-m}^2(a)\sigma^+(\varepsilon)$ is the dynamical self-energy for the Stark state $|\phi_d^{(n)}\)>$. 
Similarly, the continuous left-eigenstate has been obtained except that the delayed analytic continuation was not taken \cite{Petrosky91Physica,Yamada12PRB}.
The dressed continuous right-eigenstate $|\Phi_k^{(n)}\)>$ and the left-eigenstate $\(<\tilde\Phi_k^{(n)}|$ have the same real eigenvalues of  $z_k^{(n)}=\omega_k+n\omega$. 
The right- and left-eigenstates of the resonance states and the dressed continuous states bi-completes the decomposition of the identity in the composite space $\cal F$ as
$\hat I_{\cal F}=\sum_n \left\{ |\Phi_d^{(n)}\)>\(<\tilde\Phi_d^{(n)}|+\int dk |\Phi_k^{(n)}\)>\<\tilde\Phi_k^{(n)}|\right\} $.
This decomposition makes it possible to represent any state vector of the total system in terms of the complex spectral expansion.
By using the Fourier transform in the $\cal T$ space, the state vector at time $t$ in the space $\cal R$ is represented by
 \begin{align}\label{eq:Psit}
 |\Psi(t)\>&=e^{-iz_{d}^{(0)}t}|\Phi_{d}^{(0)}(t)\>\<\widetilde{\Phi}_{d}^{(0)}(0)|\Psi (0)\>\notag\\
 &+\int dk  e^{-i\omega_{k}t}|\Phi_{k}^{(0)}(t)\>\<\widetilde{\Phi}_{k}^{(0)}(0)||\Psi (0)\> \;.
 \end{align}

{\it HHG spectrum.}
In this work, the HHG spectrum is studied in the case where the TLS is excited from $|g\>$ to $|e\>$ at $t=0$ by a delta-function pulse, where the spectral intensity is proportional to the intensity of the excitation pulse.
In this case, the spontaneous HHG photon emission spectrum, defined as the probability of detecting a photon of frequency $\omega_k$ during the interval $t$, is obtained by $S(\omega_k,t)=\<a_k^\dagger a_k\>_t=\left|\<k|\Psi (t)\>\right|^2$ with $|\Psi(0)\>=|d\>$ \cite{GlauberBookChap,CarmichaelBook1}. 
Substituting the right- and left-eigenstates of $\hat H_{\rm F}$  in Eq.(\ref{eq:Psit}), the analytical expression for the spectral amplitude is obtained as
\begin{subequations}\label{eq:kPsit}
\begin{align}
&\<k|\Psi(t)\>=-\lambda{\cal C}_k\sum_{m=-\infty}^\infty  e^{-i z_{d}^{(m)}t}{ J_{m}(a) e^{-i m \theta} \over [\omega_k-z]^+_{z=z_d^{(m)}} }\notag \\
&+\lambda{\cal C}_k e^{-i\omega_k t}\sum_{m=-\infty}^\infty { J_{m}(a) e^{-i m \theta} \over \omega_k+i0^+-{\cal A}_m^+(\omega_k)  }  \notag\\
& +{i\over 2\pi} \lambda {\cal C}_k \sum_{n,l} \int_\Gamma d\omega' \rho(\omega') { {\cal C}_{\omega'}^2 J_l^2(a) e^{-i(\omega'-m\omega)t} \over \sum_m J_{l-m}^2(a) {\cal C}_{\omega'-m\omega}^2 \rho(\omega'-m\omega) }\notag\\
&\qquad \times {1\over\omega'-{\cal A}_l^\Gamma(\omega') }{J_{l-n}(a) e^{-i(l-n)\theta} \over \omega'+i0^+-(\omega_k+n\omega) }\\
&\equiv s_{\rm R}(k,t)+ s_{\rm C}(k,t)+ s_{\rm BR}(k,t)\;,
\end{align}
\end{subequations}
where $\rho(\omega)$ is the density of states of the continuum, and  ${\cal C}^2_kdk={\cal C}(\omega)^2\rho(\omega)d\omega$.

\begin{figure}[t]
\includegraphics[width=8cm,height=3.5cm]{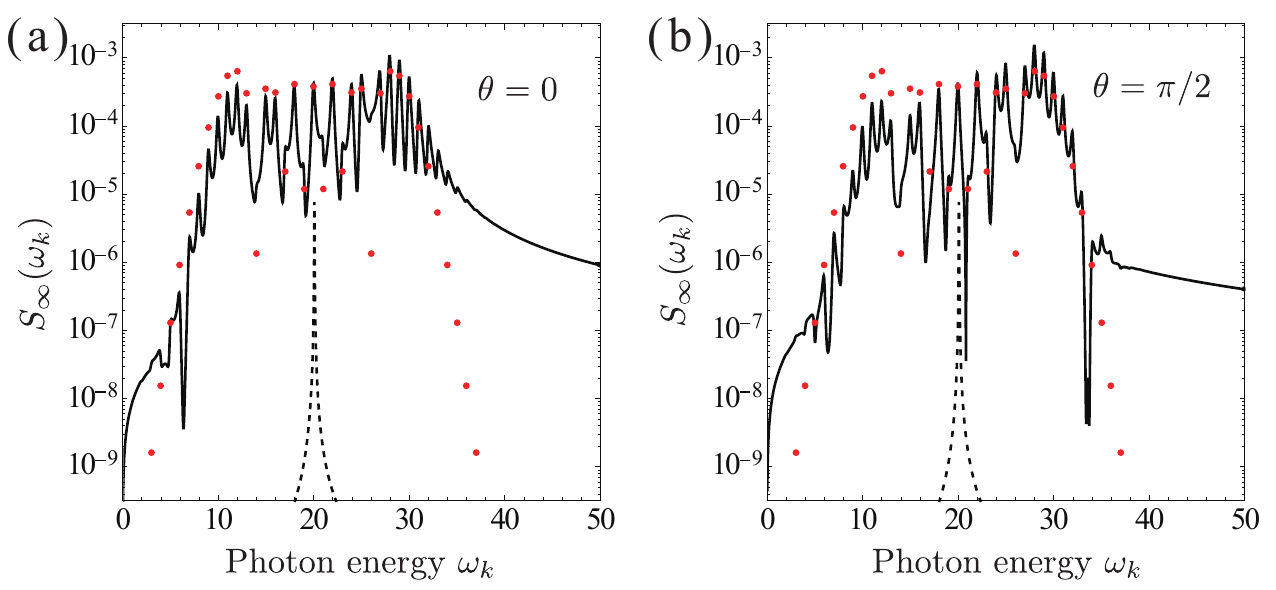}
\caption{Stationary HHG spectrum $ S_\infty(\omega_k)$ for $\Delta_0=20, a=10, \lambda=0.06$. (a) $\theta=0$ and (b) $\theta=\pi/2$. 
The fundamental spontaneous emission spectrum at $\omega_k=20$ is shown by the dashed lines.
The red marks indicate the absolute value of the Bessel's function $|J_l(a)|^2$. }
\label{fig:StatHHG}
\end{figure}

Equation (\ref{eq:kPsit}) is the principal result of this paper: The contributions of the resonance state and the dressed continuous states are analytically decomposed in the first and second terms, respectively. 
While the contribution of the resonance state decays exponentially with time, that due to the continuous state gives a stationary HHG spectrum.
The third term is attributed to the {\it branch point effect}, where the contour of the integral denoted by $\Gamma$ is taken in the different Riemann sheets at the branch point.
This term represents the non-Markovian effect, only contributing to the very short time known as {\it Zeno time}, or the very long time known as the long-time tale \cite{Petrosky01PRA}.
It is seen in the present case that, with a large amplitude of the driving laser field, the contribution of the third term is very small.

The stationary HHG spectra $S_\infty(\omega_k)\equiv\lim_{t\to \infty}S(\omega_k,t)$ is shown in Fig.\ref{fig:StatHHG} for the parameters: $\Delta_0=20, a=10, \lambda=0.06$, and $\theta=0$ in (a) and $\theta=\pi/2$ in (b).
These parameters are approximately of the same order as those in the experiments with semiconductors\cite{Ghimire11NatPhys,Schubert14NaturePhoto}.
The characteristic features of the HHG spectrum, such as the plateau and  cutoff, are well reproduced in the figures.
Overall, the absolute values of the Bessel's function $|J_m(a)|^2$ determine the spectrum intensities as shown by the red marks, so that 
the cutoff energy is approximately determined by the amplitude of the laser field $a$, and not by the intensity $a^2$, underlining the typical feature of the HHG spectrum from solids\cite{Ghimire11NatPhys,Schubert14NaturePhoto,Hohenleutner15Nature,Ndabashimiye16Nature}.

Cross terms of the different Floquet modes in $S_\infty(\omega_k)$ cause the quantum interference of the photon emissions from them.
Due to this interference effect, Fano-type dip structures appear in the plateau region as shown in Fig.\ref{fig:StatHHG}.
Since the coefficients in the summation in Eq.(\ref{eq:kPsit}) depend on the initial phase of the laser, the spectral profile of the stationary HHG spectrum is also affected, as shown in Fig.\ref{fig:StatHHG}(a) and (b).
Hence, it is possible to quantum mechanically control the HHG photon emission by changing the initial phase of the driving laser field.

\begin{figure}[t]
\includegraphics[width=8cm,height=6cm]{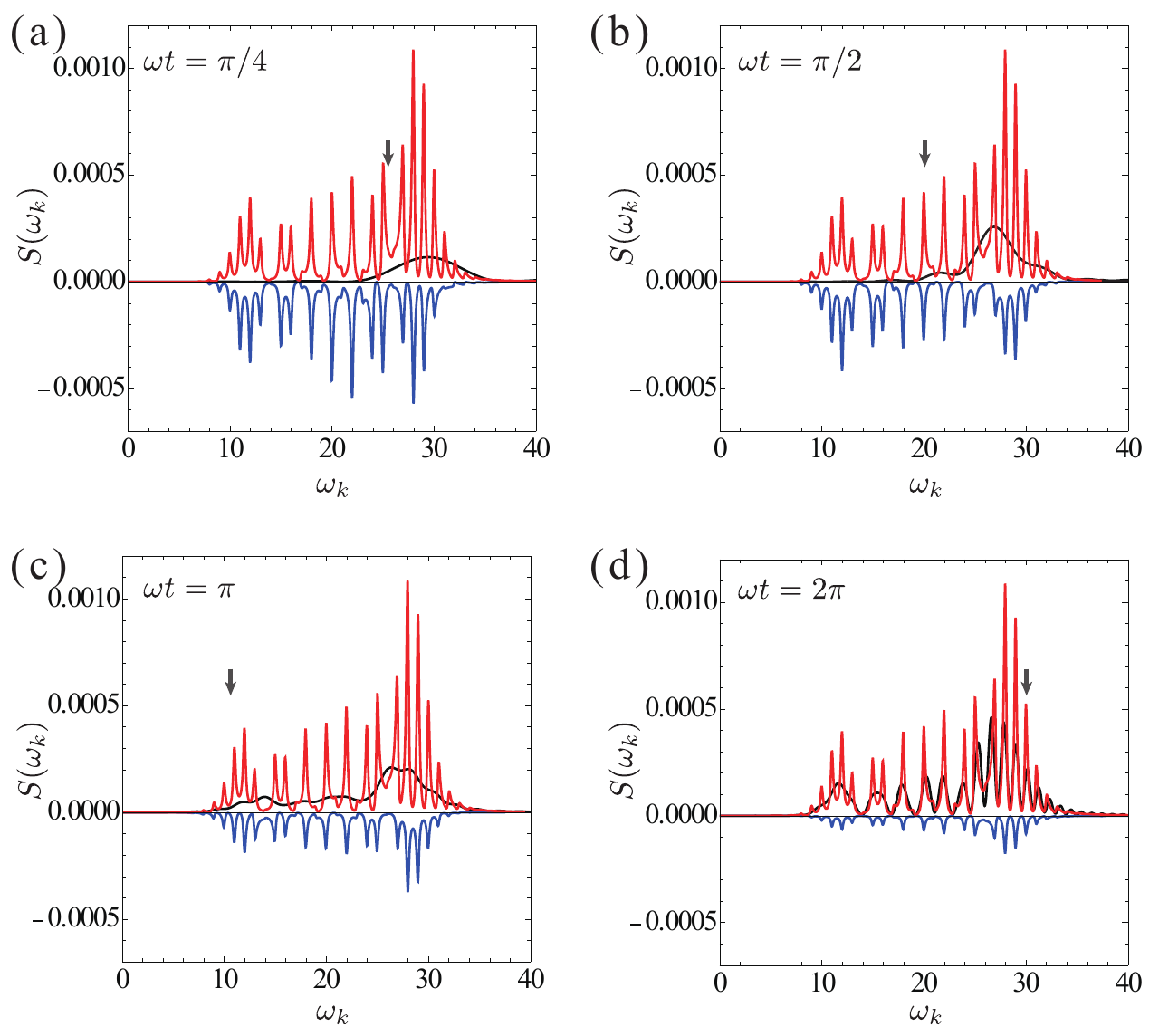}
\caption{The temporal spectral profile of HHG $S(\omega_k,t)$ (black line)  for $\omega t=\pi/4$ (a), $\pi/2$ (b), $\pi$ (c), and $2\pi$ (d), where the same parameters of Fig.\ref{fig:StatHHG}(a) are used. 
The resonance state $|s_{\rm R}(k,t)|^2$ and the dressed continuous state $(-1)|s_{\rm C}(k,t)|^2$  components are depicted by the blue and red lines, respectively.
The excited state energies $E_e(t)=E_e+A\cos(\omega t+\theta)$ are indicated by the arrows. }
\label{fig:TempHHG}
\end{figure}

Within the lifetime of an excited state, the resonance state components crucially contribute to the temporal profile of the HHG spectrum.
As seen from Eq.(\ref{eq:kPsit}) the first and second terms have opposite signs, hence the spectral amplitude cancels out at $t=0$ except for the small branch point effect.
As the resonance component decays exponentially with time, the spectral cancellation weakens, resulting in the stationary HHG spectrum.
In Fig.\ref{fig:TempHHG} the temporal change of the HHG spectrum is shown, where the components of the resonance and dressed-continuous states are separately depicted.
Although the spectral cancellation of the resonance and continuous states has been studied in configurational space domain for a simple spontaneous emission system\cite{Petrosky01PRA}, the present result is the spectral cancellation in the frequency domain under a strong driving field.

\begin{figure}[t]
\includegraphics[width=8cm,height=7.5cm]{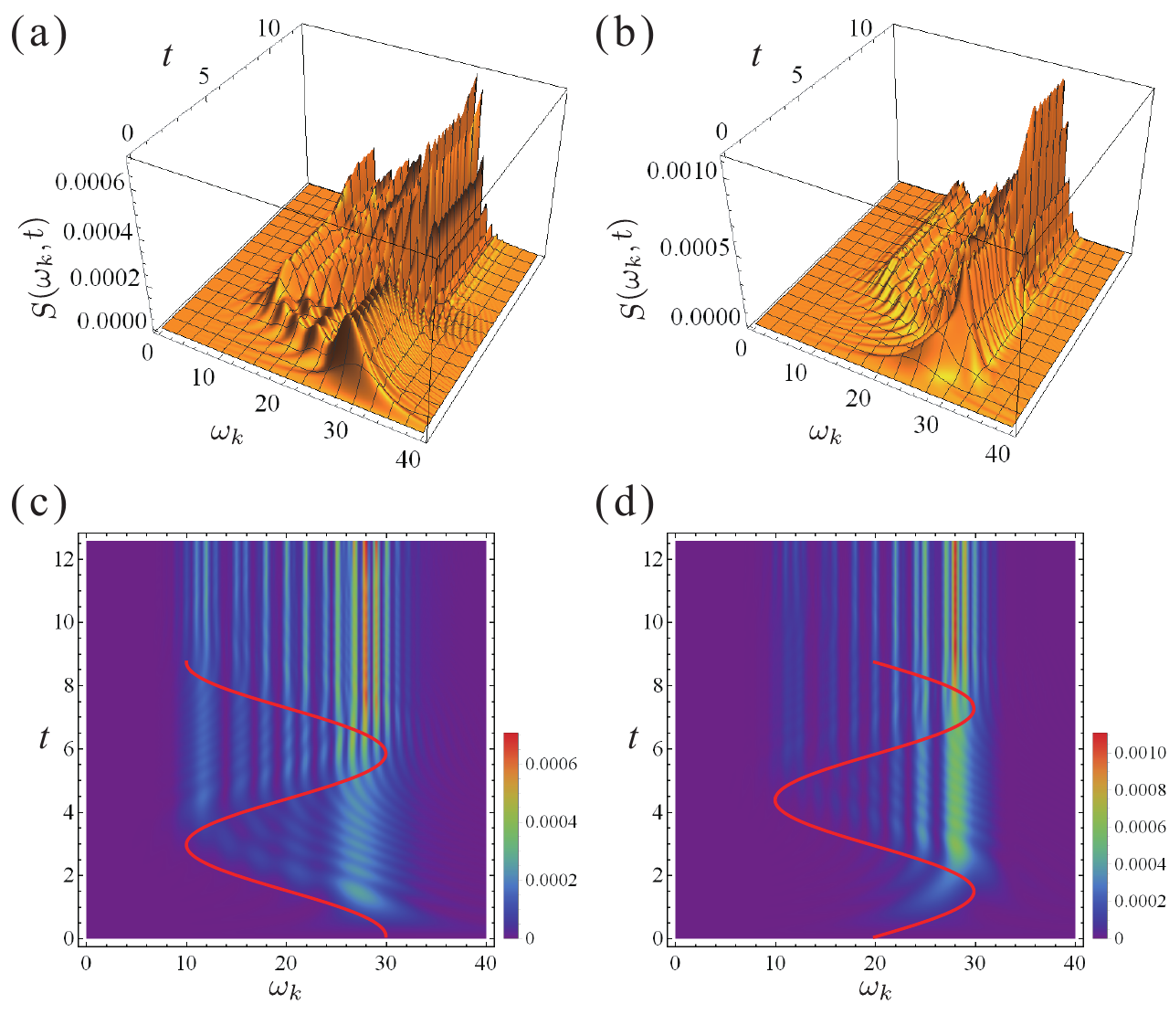}
\caption{The transient HHG spectrum for  $\theta=0$ ((a),(c) ), and $\theta=\pi/2$ ((b),(d)). The parameters are the same as Fig.\ref{fig:StatHHG}.
The red curves in the contour maps (c) and (d) indicate  $E_e(t)$.}
\label{fig:HHGmap}
\end{figure}

The resonance state components of the HHG not only reduce its intensity but also change its spectral shape due to the interference of the Floquet resonance modes, as shown by the blue curves in Fig.\ref{fig:TempHHG}, while the dressed-continuous state components retain its spectral shape.
Due to the interference of the Floquet resonance states, the peak position of the HHG spectrum adiabatically follows the temporal excited state energy $E_e(t)=E_e+A\cos(\omega t+\theta)$ as shown in Fig.\ref{fig:HHGmap}.
In time, the adiabatic behavior of the transient HHG asymptotically approaches the stationary HHG spectrum.

{\it Summary.}
In this Rapid Communication, the HHG from a TLS driven by a monochromatic phase-locked laser field has been studied in terms of complex spectral analysis for the total system.
The Floquet resonance state as well as the dressed continuous scattered radiation field were obtained as eigenstates of the total system such that the quantum correlation was maintained throughout the HHG process.
The calculated results exhibit the feature of the HHG spectrum in solids, even in the TLS system, since the two levels can be considered to correspond to the valence and conduction bands at $k=0$ states \cite{Korbman13NewJPhys,Wu16PRA}. 
Recent studies have pointed out the quantum interference between the multiband transition in the HHG process in solids \cite{Schubert14NaturePhoto,Hohenleutner15Nature}, but it has been clarified in this study, that the quantum interference of Floquet resonance states also plays an essential role in the dynamics of HHG photon emission.
The crucial role of the interference is clarified, in the transition from the initial adiabatic regime to the stationary HHG spectrum, as a result of the coherent quantum dynamics of the total system, in contrast to the phenomenological approaches.

{\it Acknowledgements.}
We are very grateful K. Kanki, S. Garmon, Y. Kayanuma, M. Domina, and N. Hatano for fruitful discussions.
This work was partially supported by JSPS KAKENHI Grants No. JP16H04003, No. JP16K05481, and No. JP17K05585.


\end{document}